\documentclass[final,leqno]{article}

\usepackage[latin1]{inputenc}
\usepackage{graphicx}
\usepackage{amsmath,amssymb,bm}

\newcommand{\bb}{\mathbf{b}}
\newcommand{\x}{\mathbf{x}}
\newcommand{\nn}{\mathbf{n}}

\newcommand{\pder}[2]{\frac{\partial #1}{\partial #2}}
\newcommand{\pders}[4]{\frac{\partial^{#3} #1}{\partial #2^{#4}}}

\newcommand{\F}{V}
\newcommand{\m}[1]{\mathrm{m}_n^{#1}}
\newcommand{\M}[1]{\mathcal{M}_n^{#1}}

\newcommand{\He}{\mathbb{H}}
\newcommand{\A}{\mathbb{A}}
\newcommand{\Mat}{\mathbb{M}}
\newcommand{\dM}{\mathrm{M}\,}
\newcommand{\dMij}{\mathrm{M}_{ij}\,}

\newcommand{\Rg}{\mathcal{R}}
\newcommand{\Null}{\mathcal{N}}
\newcommand{\cc}{\mathbf{c}}

\newcommand{\ku}{\kappa_1}
\newcommand{\kd}{\kappa_2}
\newcommand{\kt}{\kappa_3}
\newcommand{\ki}{\kappa_i}
\newcommand{\kj}{\kappa_j}

\newtheorem{theorem}{Theorem}

\begin{document}

\title{A Catastrophe-Theoretic Approach to Tricritical Points with Application to Liquid Crystals}

\author{Livio Gibelli 
\thanks{Dipartimento di Matematica, Politecnico di Milano, Piazza Leonardo da Vinci 32, 20133 Milan, Italy
({\tt livio.gibelli@polimi.it})}
\and Stefano Turzi \thanks{Department of Chemistry, University of Southampton, Southampton SO17 1BJ, United Kingdom
({\tt stefano.turzi@soton.ac.uk})}}

\maketitle

\begin{abstract}
A criterion to locate tricritical points in phase diagrams is proposed. The criterion is formulated in the framework of the Elementary Catastrophe Theory and encompasses all the existing criteria in that it applies to systems described by a generally non symmetric free energy which can depend on one or more order parameters. We show that a tricritical point is given whenever the free energy is not 4-determined. An application to smectic-C liquid crystals is briefly discussed.
\end{abstract}

\pagestyle{myheadings}
\thispagestyle{plain}
\markboth{L. GIBELLI AND S. TURZI}{A Catastrophe-Theoretic Approach to Tricritical Points}

\section{Introduction}
Despite a common belief, Catastrophe Theory (CT) can provide not only qualitative insight but also quantitative results. This is particularly true for phase transitions of systems whose free energy depends on several order parameters, such as those occurring in liquid crystals. 
Usually, CT is applied to concrete examples by identifying the elementary catastrophe associated with each transition. Calculations can then most easily and efficiently be done on this equivalent form. In doing this translation from a physical to a pure mathematical realm, one retains the intuition of the problem, but at the same time looses the ability to write explicit expressions for the physical quantities involved in the transition.\\
One of the aims of the present paper is to show how quantitative equations can be drawn from CT together with a geometric and intuitive description of their meanings when the physical conditions are recast in a diffeomorphic invariant form.

The particular case of tricritical points in 4-order parameter systems is studied. Tricritical points are points in thermodynamic phase space where a phase transition of a complex kind takes place involving the meeting of a second-order transitions with a line of first-order transitions. 
In mathematical terms, they turn out to be the points with the highest degeneracy in a butterfly $A^+_5$ catastrophe. We then apply our results to a specific example drawn from liquid crystals theory, namely smectic-C liquid crystals. A possible subsequent application is the study of tricritical points in biaxial liquid crystals. \\

This paper is organized as follows. In Section \ref{sec:CT} we introduce the mathematical background necessary to understand the rest of the paper. In particular we omit a few technical details that are important but can be easily studied form the general literature and could otherwise obscure the reading of the paper. Section \ref{sec:criterion} is devoted to develop the main result: we derive a criterion to identify tricritical points in the control space. In Section \ref{sec:simpler} we specify our main result to simpler cases which are physically important and show that our criterion reduces to the ones already known in the literature when additional ad hoc assumptions are made.
An application to smectic liquid crystals is given in Section \ref{sec:smectic} and the conclusions are drawn in Section \ref{sec:conclusions}.

\section{Mathematical background}\label{sec:CT}
In the present Section we recall the very basic features of CT that will be useful in the following. We refer the reader to the excellent introductions \cite{93gilm,99dema,93okad} as well as to the review paper \cite{78golu} for further details.

Let $V({\bf x},\boldsymbol{\lambda})$ be a real-valued smooth function with 
{\em variables} ${\bf x} = (x_{1}, x_{2}, \ldots, x_{n}) \in \mathbb{R}^{n}$ 
and {\em parameters} $\boldsymbol{\lambda} = 
 (\lambda_{1}, \lambda_{2}, \ldots, \lambda_{p}) \in \mathbb{R}^{p}$.
For each fixed $\boldsymbol{\lambda}$ one obtains a function 
$V_{\boldsymbol{\lambda}}:\mathbb{R}^{n}\rightarrow\mathbb{R}$ and we are thus
enabled to interpret $V$ as a $\lambda$-parameter
family of smooth functions from $\mathbb{R}^{n}$ to
$\mathbb{R}$. 
The function $V_{\boldsymbol{\lambda}}$ has a {\em singular point} at 
${\bf x}_{0}$ for $\boldsymbol{\lambda}=\boldsymbol{\lambda}_{0}$ if 
$\partial_{\bf x}V_{\boldsymbol{\lambda}_{0}}({\bf x}_{0})={\bf 0} 
 \; \; \forall {\bf x}$.
We assume that 
${\bf x}_{0}=\mathbf{0}$, $\boldsymbol{\lambda}_{0}={\bf 0}$ and  
$V({\bf 0},{\bf 0})={\bf 0}$, 
as the general case is deduced from this by translation.
We aim at analysing the behaviour of 
$V_{\boldsymbol{\lambda}}$ in the neighbourhood of ${\bf 0}$, as the 
parameters $\boldsymbol{\lambda}$ are varied.
Since we are interested in a local study, we lump together all functions
which coincide with $V_{\boldsymbol{\lambda}}$ near ${\bf 0}$ and call this 
set of functions the {\em germ} of $V_{\boldsymbol{\lambda}}$ at ${\bf 0}$.
For later reference, let $m_{n}$ denotes the set of germs $f(\mathbf{x})$ with 
$f(\mathbf{0})=\mathbf{0}$. 
More generally, we denote by $m_{n}^{k}$ the set of germs in $m_{n}$ such that 
all their partial derivatives of order less than $k$ vanish at ${\bf 0}$.
These powers of $m_{n}$ form a descending chain, i.e. 
$m_{n} \supseteq m_{n}^{2} \supseteq m_{n}^{3} \supseteq \ldots$. It can be shown that $\m{k}$ is an ideal generated by all monomials of homogeneous degree $k$.

The germ $V_{\boldsymbol{\lambda}}$ is said to have a 
{\em non-degenerate or Morse}
singular point at ${\bf 0}$ if its Hessian is there nonsingular, i.e.
the rank of the Hessian is $n$. 
Otherwise the singular point is {\em degenerate or non-Morse}.
At a Morse singular point, the germ  
$V_{\boldsymbol{\lambda}}$ exhibits a minimum, a maximum or a 
saddle point and is {\em structurally stable}, that is 
all the germ obtained by a small change of $\boldsymbol{\lambda}$
have the same kind of singularity.
Moreover, a theorem of Morse guarantees that there exists a coordinate
transformation which permits to write $V({\bf x},\boldsymbol{\lambda})$,
in a neighbourhood of ${\bf 0}$, as

\begin{equation}
\label{Morse}
V_{\mbox{\tiny M}} ({\bf x})
      = - \left( x_{1}^{2}+\ldots+x_{k}^{2} \right)
        + \left( x_{k+1}^{2}+\ldots+x_{n}^{2} \right)
\end{equation}
where $k$ is the index of Hessian matrix at ${\bf 0}$, 
i.e. the number of negative eigenvalues of the matrix.

At a non-Morse singular point the germ is structurally unstable 
and varying the control parameters it undergoes a catastrophe, that is
neighbouring germs may possess a different number of singular 
points with different nature.   
The fundamental example is $V_{\lambda}(x)=x^{3}-\lambda x$, which has one 
minimum when $\lambda>0$ and none when $\lambda<0$.
Non-Morse singularities can be further classified by means of corank, 
determinacy and codimension of $V_{\boldsymbol{\lambda}}$ at ${\bf 0}$.
The corank, $c$, is the number of eigenvalues of the Hessian matrix of 
$V_{\boldsymbol{\lambda}}$ which are nought.
We may equivalently think of it as the number of directions in which the 
germ is degenerate.
It can be proved that, in a suitable coordinate system,
$V({\bf x},\boldsymbol{\lambda})$ can be written as the sum of a Morse, 
$V_{\mbox{\tiny M}}$, and non-Morse part, $V_{\mbox{\tiny NM}}$, that is
 
\begin{equation}
V ({\bf x},\boldsymbol{\lambda})= 
    V_{\mbox{\tiny M}} ({\bf x}_{i}) +
    V_{\mbox{\tiny NM}} ({\bf x}_{e}, \boldsymbol{\lambda})  
\end{equation}
where ${\bf x}_{e}=\left\{x_{1},\ldots,x_{c}\right\}$ and
${\bf x}_{i}=\left\{x_{c+1},\ldots,x_{n}\right\}$ are named
{\em essential} and {\em inessential} variables, respectively. 
This result is called {\em splitting lemma} because it allows to split the
variables into two classes. Unlike the inessential variables, the
essential variables are involved in the structural instability and
the kind of catastrophes which can occur 
depends only on their number, i.e. the corank $c$ of the singularity. 
If the number of control parameters is not greater than five, 
{\em Thom's theorem} states that the non-Morse part $V_{\mbox{\tiny NM}}$ can be put into a canonical form,
named {\em elementary catastrophe}, which is given by the sum of a
germ which only depends on ${\bf x}_{e}$, said {\em catastrophe germ},
and a family of germs which 
depend on ${\bf x}_{e}$ as well as on $\boldsymbol{\lambda}$, said
{\em catastrophe perturbation}.
The latter is also named {\em unfolding}. The idea is that the
degenerate critical point of $V_{\boldsymbol{\lambda}}$ can be potentially unfolded under a perturbation into several non-degenerate critical 
points, which then appear in the nearby germs. Table \ref{tab:elemcat} gives the list of the corank one elementary catastrophes. \\
Determinacy, $k$, and codimension, $r$, of $V_{\boldsymbol{\lambda}}$ at 
${\bf 0}$ permit to determine both the catastrophe germ and germ
to which $V_{\mbox{\tiny NM}}$ is equivalent in a suitable coordinate 
system.
We define the $k-\mbox{jet}$ of the germ $f({\bf x})$, 
which we write as $j^{k} f({\bf x})$, to be the formal Taylor 
expansion up to including terms of order $k$.
A germ $f({\bf x})$ is {\em k-determined} if every germ $g({\bf x})$ 
having $j^{k} f = j^{k} g$ is diffeomorphic to $f$. 
The determinacy of a one-variable germ is trivial since it is simply given by
the first non-zero term in its Taylor expansion. In more than one dimension 
this is no longer the case. The
germ $x_{1}^{2}x_{2}$, for instance, is four-determined yet its Taylor
expansion contains only three-order terms. It is worth to notice that
a germ at a Morse singular point is two-determined. \\
Algebraic criteria for the determinacy of a germ $f({\bf x})$ are based on 
the so-called {\em Jacobian ideal}, $J[f({\bf x})]$. This is the ideal generated by $\partial f/\partial x_{i}$, i.e. 
$J[f({\bf x})]\,=\, \{ g_{1}\partial f/\partial x_{1} + \ldots + g_{n}\partial f/\partial x_{n} \}$  for arbitrary germs
$g_{i}({\bf x})$. 
By means of the Jacobian ideal, we may also define the codimension of a germ $f({\bf x})$ as the dimension of the quotient space
\begin{equation}
\mbox{cod} \left(f\right) = \mbox{dim} \left(\m{}/J[f] \right) \, .
\end{equation}
We will go over the concepts of determinacy and codimension by discussing the 
Mather's criterion in Section \ref{sec:criterion}.

\begin{table} \begin{center}
\begin{tabular}{|ll|l|c|c|c|l|} \hline
Name & & k & r & cat. germ & unfolding  \\ \hline \hline
fold & $A_{2}$ & 3 & 1 & $x^{3}$ & $a_{1}x$ \\ \hline
cusp & $\displaystyle A^{\pm}_{3}$ & 4 & 2 & $\pm x^{4}$ & $a_{1}x+a_{2}x^{2}$ \\ \hline
swallowtail & $A_{4}$ & 5 & 3 & $x^{5}$ & $a_{1}x+a_{2}x^{2}+a_{3}x^{3}$ \\ \hline
butterfly & $\displaystyle A^{\pm}_{5}$ & 6 & 4 & $\pm x^{6}$ & $a_{1}x+a_{2}x^{2}+a_{3}x^{3}+a_{4}x^{4}$ \\ \hline
wigwam & $A_{6}$ & 7 & 5 & $x^{7}$ & $a_{1}x+a_{2}x^{2}+a_{3}x^{3}+a_{4}x^{4}+a_{5}x^{5}$ \\ \hline
\end{tabular} \caption{\small Thom's elementary catastrophes of corank 1. $k$: determinacy of germs; $r$ codimension}
\label{tab:elemcat}
\end{center} \end{table}


\section{Criterion for tricritical points}\label{sec:criterion}

\begin{figure}[th] \centering
\includegraphics[width = 0.5\textwidth]{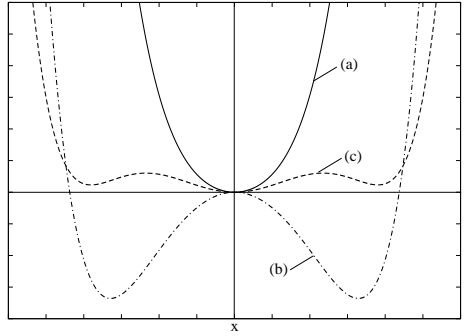}
\caption{\small Potential (\ref{eq:landau}) as the control parameters are varied. (a) $a_2>0$, $a_4>0$ ; (b) $a_2<0$, $a_4>0$ ; (c) $a_2>0$, $a_4<0$.}
\label{fig:esempio}
\end{figure}

Here we will assume that we are given a system whose phases are described by 4 order parameters (namely $x,y,z,w$) and that the equilibrium phase is attained at a minimum of a free energy potential $V(x,y,z,w)$ which is taken to be a smooth function. Moreover, we assume that $V$ also depends on $k \leq 5$ physical parameters, i.e., at the most 4 model parameters $\lambda_i$ plus the temperature $T$. For ease of notation, the physical parameters are not always explicitly indicated, but it is understood that $V$ depends on them. As a consequence, also the position of the critical points will depend on $\lambda_i$ and $T$. We will follow \cite{73grif} and define a tricritical point as a point in the phase diagram where a first order transition becomes second order. 
Our aim is to give, from a catastrophe-theoretic standpoint, a criterion that the physical parameters have to satisfy in order to identify a tricritical point in the phase diagrams of complex systems such as smectic or biaxial liquid crystals.\\
In the simplified model describing the phase of a system in terms of a single order parameter $\psi$, the free energy is usually given in terms of a Landau expansion as 
\begin{equation}\textstyle \label{eq:landau}
V\,=\,\frac{1}{6}\psi^6 + \frac{1}{4} a_4 \psi^4 + \frac{1}{2} a_2 \psi^2
\end{equation}
where $a_4, a_2$ are the control parameters. It is well known \cite{80land,73grwi} that second order phase transition points in the phase diagram are given by $a_2=0, a_4>0$ and the existence of a tricritical point is determined by $a_2=a_4=0$. This is easily understood by drawing the family of potentials (\ref{eq:landau}) as $a_2$ and $a_4$ are varied (see Figure \ref{fig:esempio}) . The first condition corresponds to the appearance of two new minima placed symmetrically with respect to $\psi=0$, while the phase $\psi=0$ becomes unstable and the transition to the new phase is therefore continuous or second-order. The second condition marks the transition to a higher degenerate potential $V$ where the phase at $\psi=0$ is still stable, therefore from there on the phase transition to a $\psi\neq 0$ is expected to be first order.\\

We would like to extend the simple 1-variable tricriticality criterion $a_2=a_4=0$ to more complex cases, preserving at the same time its valuable feature of being intuitively clear. 
Here, we have assumed the potential to be even, but this condition will be dropped in the following general treatment. A symmetric case will be considered in Section \ref{sec:simpler}.\\
Of course, it is not possible to simply say ``the fourth coefficient of the Taylor expansion at the transition is nought'' as (a) the critical points are not fixed but a slight variation of the control parameters implies a change in the equilibrium order parameters and (b) there are more than one variable involved in the transition and it is not a-priori known along which direction the bifurcation is going to take place. In \cite{86long,89long} point (a) has been addressed by the introduction of a ``master'' order parameter which is different from zero only in the ordered phase. All the remaining order parameters are supposed to depend on it. We will not need to make this assumption here.\\
At equilibrium, for given $\lambda_i$ and $T$, the order parameters solve $\nabla V(x,y,z,w;\lambda_i,T)=0$ and we denote by $P_0=(x_0(\lambda_i,T),y_0(\lambda_i,T),z_0(\lambda_i,T),w_0(\lambda_i,T))$ the critical point of our interest.
To avoid unnecessary generality and to make our study close to the application in liquid crystal theory, we make the following assumptions:
\begin{enumerate}
\item $V(x,y,z,w)$ has a corank 1 singularity at the transition. \\
This means that the hessian matrix $\He$ is singular, i.e. $\det(\He)=0$, but only one eigenvalue of $\He$ will be nought. Hence, we lose no generality in assuming that $V_{zz} \neq 0$
\item The third and fourth column of $\He$ are linearly dependent. \\
Since from previous point $\det(\He)=0$, we know that the columns of $\He$ must be linearly dependent. Here, we want to express the condition that the bifurcation will involve only the $z$ and $w$ variables. After the transition new minima will appear and if we check their location in the $(z,w)$ plane, they will move along a direction determined by the eigenvector associated with the null eigenvalue (see \cite{93gilm}). No bifurcation will involve $x$ or $y$. Therefore, another way of stating the assumption is: the eigenvector associated with the null eigenvalue is of the form $(0,0,v_3,v_4)$.
\end{enumerate}

Now CT can be straightforwardly applied. Following assumption 1, the splitting lemma allows us to study the functions with only one degenerate variable. Therefore, apart from a diffeomorphic change of variables that brings the ``physical variables'' $(x,y,z,w)$ into the ``mathematical variables''  $(\alpha_1,\alpha_2,\alpha_3,\psi)$ and the ``physical parameters'' $(\lambda_i,T)$ into the ``mathematical parameters'' $a_i$; the potential $V$ will locally be of the type: $\tilde{V}(\alpha_1,\alpha_2,\alpha_3,\psi)\,=\, \frac{1}{2} \alpha_1^2 + \frac{1}{2} \alpha_2^2+ \frac{1}{2} \alpha_3^2 + f(\psi;a_i)$, where in the list of elementary catastrophe (see Table \ref{tab:elemcat} or \cite{93gilm,93okad}) the least degenerate (lowest codimension) example of a germ which is physically acceptable (bounded from below) and allows a tricritical point, is the butterfly $A^+_5$: $f(\psi)=\frac{1}{6}\psi^6 + \frac{1}{4}a_4 \psi^4 + \frac{1}{3}a_3 \psi^3 + \frac{1}{2}a_2 \psi^2 + a_1 \psi$.
In the case of even potentials, odd powers have to be ruled out for symmetry reasons. This is equivalent to the usual expression found in physics books \cite{80land}. \\
By standard results in CT, all possible perturbations of the potential $V$ in an open set around the critical point, i.e. all the possible topologically equivalent phase diagrams are given by the above expression when the $a_i$ are varied. If, for ease of calculation, we assume that $V$ has a critical point at $\psi=0$, then $a_1=0$. It can now be shown that also in the multi-variable case, the phase $\psi=0$ looses stability when $a_2=0$ and a second order transition happens at $\psi=0$ when $a_2=a_3=0, a_4>0$. The presence of a term $a_3 \neq 0$ in general signals a first order transition to a phase with $\psi \neq 0$ and therefore makes the whole study of the transitions much richer (and complex). Anyway, since we are here only interested in tricritical points, i.e. limit points where a transition pass from second order to first order, we do not need to exploit such complexity but can limit ourselves to the regions of the control space where second order transitions are possible. It is therefore recognised that the tricritical point in $A^+_5$ is given by what is konwn as the germ of the catastrophe, the point with maximum degeneracy (see \cite{80land,93gilm,73grwi,86long}), given by
\begin{equation}
a_1=a_2=a_3=a_4=0.
\label{eq:tri0}
\end{equation}

The condition (\ref{eq:tri0}) needs to be translated in terms of the physical parameters $(\lambda_i,T)$ if it has to be directly applied to a real world example, i.e., to a potential $V(x,y,z,w)$. This translation can be gained if we use the geometric language of CT to recast (\ref{eq:tri0}) in a intrinsic form, invariant over change of variables: (a) when the control parameters belong to the bifurcation set, then $V(x,y,z,w)$ is \emph{not} 2-determined and (b) when they belong to the tricritical set, $V(x,y,z,w)$ is \emph{not} 4-determined. In particular the necessary condition (b) for tricriticality is equivalent to (\ref{eq:tri0}) and becomes also sufficient when it is not trivial, i.e., in the case of least codimension (butterfly $A^+_5$ catastrophe) considered above. \\
We thus arrive at the natural multi-variable extension of the criterion (\ref{eq:tri0}) on tricritcality: \emph{a necessary condition for the control parameters $(\lambda_i,T)$ to belong to the tricritical set is that the free-energy function $V$ is not 4-determined at the critical point $P_0(\lambda_i,T)$}. \\
The effectiveness of this approach lies in the powerful tools that are now at hand. In particular, Mather \cite{99dema,78golu} has given a completely algebraic procedure to calculate the determinacy of a function.\\

\begin{theorem}[Mather] Let $V$ be a germ and let $r>0$ be an integer. If 
\begin{equation} \m{r+1} \subset \m{2}J[V] + \m{r+2} \, , \label{eq:mather} \end{equation} 
then $V$ is $r$-determined. 
\end{theorem}

As usual, the sum and product of two ideals $I$ and $J$ are the two ideals defined by $I+J:=\left\{i+j\,:\,i \in I, j\in J \right\}$ and $IJ:=\left\{ij\,:\,i \in I, j\in J \right\}$. If $I$ is generated by $\{i_a\}$ and $J$ is generated by $\{j_b\}$, then $IJ$ is generated by the products $i_a j_b$. Thus, we can define the powers $I^2, I^3, \ldots$ of an ideal $I$ taking the powers of its generators. The ideals $\m{r}$ are indeed defined in this way as powers of $\m{}$. \\
A condition stronger than (\ref{eq:mather}), yet easier to check is 
\begin{equation}
\m{r-1} \subset J[V] + \m{r} \label{eq:forte2} \, ,
\end{equation}
which follows form the properties of the ideals involved \cite{99dema}.
Then, if $V$ is not $4$-determined the two conditions (\ref{eq:mather}) and (\ref{eq:forte2}) with $r=2,3,4$ have to be false. We will now translate the stronger expression (\ref{eq:forte2}) from the abstract form into an algorithmic form and apply it to our case. As usual in CT, in the abstract description we assume with no loss of generality that the Taylor expansions are performed in the origin. We will then express the results in terms of the potential $V$ evaluated in the critical point $P_0$ (which is not necessarily the origin).\\

Let us define $\M{r}=j^r \m{r}$ the $r$-jet linear space of the germs in $\m{r}$. It contains all the linear combinations of the homogeneous monomials of degree $r$. 
To make our argument more concrete, we first concentrate on the case $r=3$ (the $r=2$ case simply translates the bifurcation condition). Thus, we look for a necessary condition that non 3-determined functions must satisfy. The case $r=4$, which renders explicit the non 4-determinacy condition, is conceptually analogous and is obtained in exactly the same way.
Taking the $(r-1)$-jet of (\ref{eq:forte2}) with $r=3$ we have
\begin{equation}
\M{2} \subset j^2 \left\{J[V]\right\} \label{eq:forte2_2}
\end{equation}
which of course needs to be false if $V$ is \emph{not} 3-determined. \\
It is now easy to build a set of generators for $j^2 \left\{J[V]\right\}$. It will be sufficient to give an example in the 1 variable case. Multi-variable case is conceptually analogous but simply involve cumbersome notations.
Elements of $J[V]$ are of the type $g(x) \pder{V}{x}$, with $g(x)$ a generic germ. We now suppose to expand $g(x)$ in Taylor series up to the second order so that the remainder $R(x)$ will be in $\m{3}$. Taking the 2-jet we have ($g_j \in \mathbb{R}$)
\begin{align}
& \textstyle j^2 \left\{g(x) \pder{V}{x} \right\}\,=\,j^2 \left\{\left(g_0 + g_1 x + g_2 x^2 + 
 R(x)\right) \pder{V}{x} \right\} \\
& \textstyle =\,g_0 j^2\left\{\pder{V}{x} \right\} + g_1 j^2\left\{x \pder{V}{x} \right\} + 
g_2 j^2 \left\{x^2 \pder{V}{x}\right\}\, . \nonumber
\end{align}

Therefore $j^2 \left\{J[V]\right\}$ is generated by all linear combinations of the polynomials of the type $q_{ij}(\x)\,=\,j^2\left\{p_i(\x) \pder{V}{x_j} \right\}$ where $p_i(\x)$ are all the monomials of degree $\leq 2$.
Following \cite{93gilm}, we are now able to state Mather's theorem (in its strongest version (\ref{eq:forte2})) in algorithmic form.\\
\begin{enumerate}
\item Calculate all the polynomials $q_{ij}(\x)\,=\,j^{r-1}\left\{p_i(\x) \pder{V}{x_j} \right\}$, where $p_i(\x)$ are all the multivariate monomials in $n$ variables, of degree from $0$ to $r-1$.
\item If all the monomials of degree $r-1$ can be obtained as linear combinations of the $q_{ij}(\x)$, then the potential $V$ is $r$-determined.\\
\end{enumerate}
In our case ($n=4$), when $r=3$ there are 10 monomials of degree 2 and 20 polynomials $q_{ij}(\x)$ whereas when $r=4$ there are 20 monomials of degree 3 and 60 polynomials $q_{ij}(\x)$. The number of the polynomials has been already reduced considering the fact that we take the jet in a critical point of $V$. However, further reductions are possible if we assume from the beginning that $V$ is not 2-determined which implies that $\det \He=0$ (i.e. we want to study the transition). The number of zero terms is greatly enhanced when symmetry conditions are further assumed on $V$ (see Section \ref{sec:smectic}).\\
The monomials of degree from 0 to 2 form a basis for both $\M{2}$ and $j^2 \left\{J[V]\right\}$, and thus the problem of 3-determinacy is  reduced with respect to this basis to the solution of a set of linear systems $\A\cc_{\gamma}\,=\,\bb_{\gamma}$, where the matrix $\A$ is built from the polynomials $q_{ij}$, and $\gamma$ is an index that ranges only on the monomials of maximum degree.

We assume here that we have chosen a degree term ordering within the multivariate monomials of degree from 0 to $r-1$. The monomials of higher degree will be placed last with respect to this ordering.
If there are $N_p$ monomials $p_i$ and $N_q$ polynomials $q_{ij}$, the matrix $\A$ and the vectors $\bb_{\gamma}, \cc_{\gamma}$ are defined as follows: 
\begin{itemize}
\item $\A \in M_{(N_p,N_q)}(\mathbb{R})$ is such that the entry in row $s$ and column $t$ is the coefficient of the $s^{th}$ monomial $p_s(\x)$ in the $t^{th}$ polynomial $q_{ij}(\x)$. Therefore, along the columns of $\A$ we can read all the coefficients of the polynomial $q_{ij}(\x)$, the $s^\mathrm{th}$-row being associated with the monomial $p_s(\x)$, 
\item $\bb_{\gamma}$ are column vectors in $M_{(N_p,1)}(\mathbb{R})$ with all first zero elements and only a 1 in $\gamma$ position to represent one of the higher degree monomials. The index $\gamma$ will range in order to solve a linear system for each of such a monomial (i.e., $x^i y^j z^h w^k$ such that $i,j,h,k \geq 0$ and $i+j+h+k = r-1$).
\item $\cc_{\gamma}$ are the vectors in $M_{(N_q,1)}(\mathbb{R})$ containing the unknown coefficients that give the linear dependence of the monomial $p_{\gamma}$ in terms of the polynomials $q_{ij}$.
\end{itemize}
Here, $M_{(h,k)}(\mathbb{R})$ is the space of real matrices with $h$ rows and $k$ columns.\\

A necessary condition for non 3-determinacy is therefore that not all such linear systems can be solved. Since it is well known from linear algebra that $\bb_{\gamma} \in \Rg(\A) \Leftrightarrow \bb_{\gamma}\in \left(\Null(\A^T)\right)^{\bot}$, this means that our criterion on tricriticality requires that there exists a $\bb_0 \in \Null(\A^T)$ such that $\bb_0 \cdot \bb_{\gamma} \neq 0$ for at least one $\bb_{\gamma}$. Here, $\Rg(\A)$ is the range of $\A$ and $\Null(\A)$ is the null space of $\A$. So we are now induced to solve the single system $\A^T \bb_0 = \mathbf{0}$ and see what condition on the entries of $\A$ can be imposed in order to get a non-trivial solution. In such a way we are simply trying to express one of the last rows of $\A$ (representing the monomials $p_{\gamma}$) as a linear combination of all the other rows.
This problem can be solved easily, only the dimensions of the matrix $\A$ suggests that it is wise to perform calculations with the aid of a symbolic computing software.
The non 4-determinacy is given by retracing the same algorithm with $r=4$. The number of polynomials involved is obviously raised and therefore the computations increase in complexity, but still a condition can be found.\\

Let us define the matrix $\Mat$ as the 3 by 3 upper-left submatrix of $\He$. It comprises the entries of $\He$ that are derivatives of $V$ with respect to the variables $x$, $y$ and $z$ only. Let $\dM\,=\,\det(\Mat)$ and $\dMij$ the $(i,j)$-minor of $\Mat$, i.e., the determinant of the 2 by 2 matrix formed by removing from $\Mat$ its $i^{th}$ row and $j^{th}$ column. \\
After some lengthy but easy algebra, we finally arrive at the conditions that a non 4-determined potential must satisfy.
\begin{enumerate}
\item $V$ is not 2-determined (and therefore a bifurcation occurs) if $\det(\He)\,=\,0$. By virtue of the assumptions on $V$ it can be shown that the above equation is equivalent to the requirement that $\det(\He_2)=0$, where $\He_2$ is the 2 by 2 lower right submatrix of $\He$.
Therefore $V$ is not 2-determined if
\begin{equation}\mathrm{D}^2 V\,=\,0 \, . \label{eq:nosym_no2det}\end{equation}
\item $V$ is not 3-determined if it is not 2-determined and 
\begin{equation}
\mathrm{D}^3 V\,=\,0 \, . \label{eq:nosym_no3det}
\end{equation}
\item $V$ is not 4-determined if it is not 3-determined and 
\begin{equation}
\dM \,\mathrm{D}^4 V - 3\sum_{i,j=1}^{3}(-1)^{i+j}\,\dMij \ki \kj \,=\,0 \, . \label{eq:nosym_no4det}
\end{equation}
\end{enumerate}
where
\begin{align}
\alpha\,&\textstyle =\, V_{zw}/V_{zz} \, ,\\
\mathrm{D}^2 V\, & =\,\alpha^2 V_{zz} - 2 \alpha V_{zw} + V_{ww} \, ,\\
\mathrm{D}^3 V\, & =\,\alpha^3 V_{zzz} - 3 \alpha^2 V_{zzw} + 3 \alpha V_{zww} - V_{www} \, ,\\
\mathrm{D}^4 V\, & =\,\alpha^4 V_{zzzz}  - 4 \alpha^3 V_{zzzw} + 6 \alpha^2 V_{zzww} - 4 \alpha V_{zwww} + V_{wwww} \, ,\\
\ku\,&=\, \alpha^2 V_{xzz}-2\alpha V_{xzw} + V_{xww} \, ,\\
\kd\,&=\, \alpha^2 V_{yzz}-2\alpha V_{yzw} + V_{yww} \, ,\\
\kt\,&=\, \alpha^2 V_{zzz}-2\alpha V_{zzw} + V_{zww}.
\end{align}
A subscript denotes differentiation with respect to the indicated variables and all the expressions are evaluated at the critical point $P_0(\lambda_i,T)$. We recall that according to our assumption 1 above, it is $V_{zz} \neq 0$. \\

In conclusion, (\ref{eq:nosym_no2det})-(\ref{eq:nosym_no4det}) are three generally independent equations in the control parameters (variables are evaluated at the critical point). They together define the set of control parameters at which a tricritical point occurs in the phase diagram. More precisely, if we have $k$ control parameters they define a $k$-3 dimensional locus of tricritical points.

\section{Reduction to simpler cases}\label{sec:simpler}
There are important simplifications to the general result that suit many physical applications. We present them here, adding more constraints to the potential $V$ as we go on in the Section. \\
First, we assume that $V$ satisfy an even symmetry with respect to a simultaneous change of sign $(z,w) \mapsto (-z,-w)$
\begin{equation}
V(x,y,-z,-w)\,=\,V(x,y,z,w). \label{eq:even}
\end{equation}
This of course implies that at the critical point
\begin{equation} 
\pders{V}{x^p \partial y^q \partial z^r \partial w}{p+q+r+s}{s}\,=\,0 \, ,\qquad \text{if } r+s \text{ is odd}.
\end{equation}

The equilibria with $z=w=0$ are then the natural candidates for phases from which a second order transition could develop. Therefore, we assume the critical point under study is $P_0=(x_0(\lambda_i,T),y_0(\lambda_i,T),0,0)$ which, as before, generally depends on the control parameters $(\lambda_i,T)$. Moreover, the hessian matrix will be block diagonal:
\[\He\,=\,\begin{pmatrix} V_{xx} & V_{xy} & 0 & 0\\
V_{yx} & V_{yy} & 0 & 0 \\
0 & 0 & V_{zz} & V_{zw} \\
0 & 0 & V_{wz} & V_{ww}
\end{pmatrix}\,=\,
\begin{pmatrix}
\He_{1} & 0 \\
0 & \He_{2}\\
\end{pmatrix}.\]
As in the general case, the bifurcation will involve only the $z$ and $w$ variables. Explicitly, in terms of the hessian matrix, we have that $\He_{1}$ is positive definite ($\det(\He_{1})>0$ and $V_{xx}>0$) and $\He_{2}$ has a null eigenvalue at the transition (and only one since we are assuming a corank 1 singularity).\\
Furthermore, this symmetry directly implies that $V$ is not 3-determined and (\ref{eq:nosym_no3det}) is automatically satisfied.
This condition (\ref{eq:even}) is found in many physical applications where a second order transition occurs. Therefore, the reduced equations are worthy of being stated explicitly
\begin{equation}
V_{zz}V_{ww} - V_{zw}^2\,=\,\det \He_{2}\,=\,0 \, ,\label{eq:non2det} 
\end{equation}
\begin{equation}
\left( V_{xx}V_{yy} - V_{xy}^2 \right)  \,\mathrm{D}^4 V - 3 \left(\ku^2 \,V_{yy} - 2 \ku\kd \,V_{xy} + \kd^2 \,V_{xx}\right)\,=\,0 \, .\label{eq:non4det}
\end{equation}

The choice of 4 order parameters has been mainly motivated by the desire to apply our criterion to existing Landau free energy expansions in liquid crystals theory, in particular to smectic or biaxial liquid crystals. However, many simplifications or reduced Landau expansions can be given where the phase transitions are well described by fewer order parameters. 
Of course, (\ref{eq:non2det}) and (\ref{eq:non4det}) ought to reduce to the equations already existing in literature when we further simplify the free energy $V$. 

For instance, in the case of two order parameters, we can declare 2 out of 4 order parameters to be ``dummy variables''. This means that a minimization with respect to this variables would always yield a constant value (which can be assumed to be zero) and no bifurcation in these variables can occur. 

The free energy will therefore be
\begin{equation}\textstyle
V(x,y,z,w)\,=\,\frac{1}{2}y^2 + \frac{1}{2}z^2 + g(x,w)
\end{equation}
where, in agreement with our previous assumption on $V$, $g(x,-w)=g(x,w)$ and at the critical point $\pders{g}{x}{2}{2}>0$. We remark here that $g(x,w)$ depends also on the control parameters, even if not explicitly indicated.
Equations (\ref{eq:non2det}) and (\ref{eq:non4det}) yield
\begin{align} 
& g_{ww}\,=\,0 \, ,\label{eq:virga1} \\
& g_{xx} g_{wwww} - 3 g_{xww}^2\,=\,0 \, ,\label{eq:virga2}
\end{align}
where again the derivatives have to be evaluated at the critical point.
The same result is to be found in \cite{05dmvi,05dmvi2} (with $x=S$, $y=S'$, $z=T$ and $w=T'$) where it is applied to biaxial liquid crystals in the fairly general yet simplified case of a null model parameter. Indeed an extension of that result has been one of the purposes of the present work and in fact (\ref{eq:non2det}) and (\ref{eq:non4det}) can in principle give the locus of tricritical points in the phase diagram of biaxial liquid crystals in their general treatment employing 4 order parameters. 
Unluckily, it is not at the present time possible to pursue this calculations any further because a reliable Landau expansion of the free energy, using 4 order parameters, is not to our best knowledge available yet. We are anyway aware that some work is being carried on by few research groups \cite{sluck} and therefore we intend to perform a study of the tricritical locus in biaxial liquid crystals in a subsequent paper when these results are published. On the other hand, we could in principle apply (\ref{eq:non2det}) and (\ref{eq:non4det}) to a non polynomial expression, such as the integral representation of the free energy given by a mean field model \cite{03sovidu,05dmvi,05dmvi2,05dmrovi,07dmbivi}. Yet, this could be done only at the cost of great computational efforts and employing numerical approximations. 

As a last example we consider a system which is described just by one order parameter $w$. The free energy can be given the form
\begin{equation}\textstyle
V(x,y,z,w)\,=\,\frac{1}{2}x^2 + \frac{1}{2}y^2 + \frac{1}{2}z^2 + f(w)
\end{equation}
where $f(-w)=f(w)$.
A substitution in (\ref{eq:non2det}) and (\ref{eq:non4det}) yields
\begin{equation}
f_{ww}\,=\,f_{wwww}\,=\,0 \, ,
\end{equation}
in agreement with the standard condition (\ref{eq:tri0}).\\
Of course, when necessary one could gather other possible reduced equations according to various combinations of dummy and non-dummy variables. The examples given here have the immediate advantage of reproducing the structure of the Landau expansions usually found in the physical literature.

\section{Application to smectic liquid crystals}\label{sec:smectic}
In this Section we apply our result to the case of smectic-C liquid crystals, where a tricritical point along the Sm-A/Sm-C transition line has been reported \cite{82huan,87gahu,88shra,95brfu,08damu}. We refer to \cite{07bicate,08damu} and bibliography therein for the physical details and to \cite{08damu} for the expression of the Landau expansion we are going to use.
Nematic liquid crystals \cite{95dgpr,94virga} are formed by elongated molecules with a cylindrical symmetry. When the thermal motion is not dominant (at sufficiently low temperature), they tend to align themselves along a common direction, usually identified by a unit vector $\nn$, called the director and form a nematic (N) phase. The degree of orientation is measured by an order parameter $s$ (where $s=0$ and $s=1$ correspond to no order and to perfect order, respectively). \\
\newcommand{\cmpuno}{B7}
\newcommand{\cmpdue}{\overline{1}\overline{0}\text{O}\overline{4}}
The smectic phase is established when an additional positional order is superimposed to the orientational order: the centre of mass of the molecules are organized in layers. The layers are described by a periodic modulation of amplitude $\rho$ in the mass density of molecules in the liquid. 
An additional order parameter $q$ gives the inverse of the distance of two subsequent layers: $d=2\pi / q$. 
When the director $\nn$ is normal to the layers, the phase is called smectic-A (Sm-A). The onset of a smectic-C (Sm-C) phase is described by a non-zero angle $\theta$ between the director and the normal to the layers. We will use $w=\sin\theta$ as the order parameter for the Sm-C phase. \\
It is known that some liquid crystal compounds show a first order Sm-A/Sm-C transition (for instance \cmpuno) while other undergo a second order transition (for instance $\cmpdue$). We follow \cite{07bicate} and study a binary mixture of two such compounds. We imagine to increase the concentration $\xi$ of one of the compounds and alter the nature of the transition. It is then reached a concentration value where the phase diagram shows a tricritical point. In a homogeneous state, the polynomial expression for the free energy density of the binary mixture is given by expression (2.5) of \cite{08damu}:
\newcommand{\s}{s_{+}}
\newcommand{\rp}{\rho_{+}}
\newcommand{\q}{q_{+}}
\newcommand{\ee}{\hat{e}}
\newcommand{\ff}{\hat{f}}
\newcommand{\al}{\hat{\alpha}}
\begin{align} 
\F\,=&\,\label{eq:free_energy} \textstyle 
\frac{1}{2}a s^2 - \frac{1}{3} b s^3 + \frac{1}{4}c s^4 + \frac{1}{2}\al \rho^2 + \frac{1}{4} \beta \rho^4
+ \frac{1}{2} \delta  s \, \rho^2 + \frac{1}{2} \gamma  s^2 \rho^2 \\
+&\, \textstyle  \frac{1}{2} d_1 \rho ^2 q^2 + \frac{1}{2} d_2 \rho ^2 q^4
+ \frac{1}{2} e s\,\rho^2 q^2 \left(2 - 3 w^2 \right) + \frac{1}{2} h s^2 \rho^2 q^4 \left(2 - 3 w^2 \right)^2  \nonumber \\
+& \, \textstyle \frac{1}{2} \eta s^2 \xi + \frac{1}{3} \omega s^3 \xi + \frac{1}{2} \lambda \rho^2 \xi + \frac{1}{2} E \xi^2\nonumber
\end{align}
Here we depart slightly from the notation found in \cite{08damu} not to generate confusion with the symbols we have already adopted in previous Sections with different meaning. The notation in \cite{08damu} is fully restored when we put $\rho=\psi_0$, $\al=\alpha$ and $\xi=x$. \\
The order parameters are: $s$, $\rho$, $q$ and $w$. They identify the various phases whenever assume a non-zero value. More precisely:
isotropic phase (I) is $s=\rho=q=w=0$; nematic phase (N) is $s \neq 0, $ and $\rho=q=w=0$; smectic-A phase (Sm-A) is $s \neq 0$, $\rho \neq 0$ $q \neq 0$ and $w=0$; smectic-C (Sm-C) phase is $s \neq 0$, $\rho \neq 0$, $q \neq 0$ and $w\neq 0$. 
At a first sight one might expect the Sm-A/Sm-C transition to be only second order because the expansion is even with respect to the order parameter $w$ and has been truncated at a point where $w$ appears only to the fourth power. It must be noticed, however, that a first order transition is possible due to the coupling between $w$ and the other order parameters. One way to see this, is to think $w$ as a master order parameter and use the equilibrium equation $\F_{s}=0$, $\F_{\rho}=0$ and $\F_{q}=0$ to express $s$, $\rho$ and $q$ as functions of $w$. It is then apparent that the various terms involving a coupling between $s$, $\rho$, $q$ and $w$ will bring higher terms in $w$ into consideration, therefore giving an intuitive justification for the presence of a tricritical point.

Following the ideas of McMillan and de Gennes \cite{89long} about the character of the N/Sm-A phase transitions, something can be argued about the transition temperatures at the tricritical point.
It is well known (as often said, due to presence of a cubic term) that $s$ undergoes a first order transition I/N at the transition temperature $T_{NI}$.  If $T_{AC}$ is the temperature at which the Sm-A/Sm-C transition takes place, close to the nematic phase ($T_{AC}/T_{NI}\approx 1$) the fluctuations of orientations are large, and as a result the character of the transition is determined by the nematic order parameter $s$, i.e., the phase transition is first order. For low transition temperature ($T_{AC}/T_{NI}\approx 0$), the nematic order is nearly saturated and thus the character of the transition is driven by $w$ and will be second order. 
Therefore, the tricritical point along the Sm-A/Sm-C transition line is expected to be in a region of the control space where $T_{AC}<T_{NA}<T_{NI}$ but $T_{AC} \approx T_{NA} \approx T_{NI}$.\\

Suppose now that we want to study the nature of the transitions from a Sm-A to a Sm-C phase. Minimization of (\ref{eq:free_energy}) with respect to $s$, $\rho$, $q$ and $w$ yields the equilibrium values in terms of the control parameters. We do not explicitly write the expressions here, simply write the equilibrium order parameters in the Sm-A phase as $s=\s>0$, $\rho=\rp>0$, $q=\q>0$ and $w=0$. With the identification $x=s$, $y=\rho$, $z=q$ and $w=w$, we can use the (\ref{eq:nosym_no2det})-(\ref{eq:nosym_no4det}) to find the conditions that the control parameters must satisfy in order to go from a second order to a first order Sm-A/Sm-C phase transition. \\
The bifurcation condition (\ref{eq:nosym_no2det}) yields 
\begin{equation} \label{eq:bifu2}
e= -4 h \,\s \, \q^2 \, ,
\end{equation}
while (\ref{eq:nosym_no3det}) is trivially satisfied.\\
Equation (\ref{eq:bifu2}) is not sufficient to guarantee that a transition takes place, as the transition might be first order. The limit point of second to first order transition is yielded by (\ref{eq:nosym_no4det}) which, after a little algebra, reads
\newcommand{\fxx}{\F_{ss}^0}
\newcommand{\fyy}{\F_{\rho \rho}^0}
\newcommand{\fzz}{\F_{q q}^0}
\newcommand{\fxy}{\F_{s \rho}^0}
\newcommand{\fyz}{\F_{\rho q}^0}
\newcommand{\fxz}{\F_{s q}^0}
\begin{align} \label{eq:tri}
\Big(\fxx \fyy & - (\fxy)^2 \Big) \left( \fzz - 16 h \s^2 \rp^2 q^2 \right) \\
& - \fxx (\fyz)^2 - (\fxz)^2 \fyy + 2 \fxy \fxz \fyz \nonumber \\
& - 4 \Big(\fyy \fzz - (\fyz)^2 \Big) h \rp^2 \q^4 + 16 (\fxz \fyy - \fxy \fyz) h \s \rp^2 \q^3 \,=\,0 \nonumber \, ,
\end{align}
where
\begin{align}
\fxx\,=&\, a - 2 b \s + 3 c \s^2 + \gamma\rp^2 + 4 h \rp^2 \q^4 + \eta\xi + 2 \omega \s \xi \, , \\
\fxy\,=&\, (\delta + 2 \gamma \s)  \rp  \, , \\
\fxz\,=&\, 8 h \s \rp ^2 \q^3 \, , \\
\fyy\,=&\, \al +\lambda + \delta \s + \gamma \s^2 +3 \beta\rp^2 + d_1 \q^2 + d_2 \q^4- 4 h \s^2 \q^4 \, , \\
\fyz\,=&\, (2 d_1 + 4 d_2 \q^3) \rp \, , \\
\fzz\,=&\, (d_1 + 6 d_2 \q^2 + 16 h \q^2 \s^2) \rp^2\, , 
\end{align}
are the entries of the hessian matrix evaluated at the critical point $(\s,\rp,\q,0)$ and (\ref{eq:bifu2}) has been used. \\
Equation (\ref{eq:tri}) can now be solved with respect to any of the control parameters (paying attention that also the equilibrium values $\s$, $\rp$ and $\q$ depend on the control parameters). Usually in Landau expansions it is assumed that only the coefficients of second order terms depend on temperature, linearly. This is usually the first step towards a physical interpretation of (\ref{eq:tri}). 
For instance, we can take $a\,=\, a_0 (T-T_1^{*})$ and $\al\,=\,\al_{0}(T-T_2^{*})$, 
where $T$ is the temperature, $T_1^*$ is the nematic supercooling temperature and $T_2^*$ is the virtual (second order) transition temperature into the smectic phase. At $T_2^*$ the solution $s=\s$, $\rho=0$ looses stability. The tricritical temperature is then given by solving (\ref{eq:tri}) with respect to the temperature $T$. It is remarkable that the same results found in \cite{08damu} after quite lengthy and tedious calculations, are given directly by (\ref{eq:tri}). In particular, (\ref{eq:tri}) is trivially satisfied when we substitute all the expressions that lead to the tricritical temperature (2.23) and the tricritical concentration (2.24) in \cite{08damu}.
However, physical significance must be given to the control parameters, for example by means of a comparison with experimental results, before physical conclusions could be drawn. This is not in general an easy task and it is outside the scopes of the present paper. \\
We must note that in general (\ref{eq:tri}) does not guarantee that a tricritical point is found, for the simple fact that the critical point we are exploring might not be the absolute minimum of the free energy. Our analysis is local. The event that the non 4-determinacy happens in a local minimum which is not an absolute minimum must be studied by means of other global techniques and a more comprehensive description of the whole phase diagram is needed.\\
Finally, we would like here to underline that our theory can in principle be applied to more complex potentials, such as a free energy functional derived by a mean-field model (see for instance \cite{05dmvi,05dmvi2} for example of applications of (\ref{eq:virga1}),(\ref{eq:virga2}) to Sm-A and biaxial liquid crystals).

\section{Conclusions} \label{sec:conclusions}
General criteria usually found in literature to locate tricritical points in the phase diagram loose their elementary interpretation when applied to complex systems such as smectic or biaxial liquid crystals. Moreover, they might not be applicable to these materials when the full generality on the number of physical parameters is allowed.\\
We have shown that rephrasing the simple and intuitive criterion employed for single order parameter systems using the invariant language of CT, allows an immediate extension to the multi order parameter cases. Therefore it is possible and often easy to explicitly write the equations that identify the tricritical points for instance in the rich and complex scenario of smectic. This extension was in fact the main aim of our work. We believe that also the interesting scenario of biaxial liquid crystals, within the general description which comprises four order parameters, can be addressed with the results described in the present paper. Our criterion can thus help in giving an analytic answer to the final question posed in \cite{05dmvi}. There the authors raise the issue whether the predicted tricritical point at the direct transition between the isotropic and the biaxial phases persists and could possibly extend when all physical parameters (and consequently all four order parameters) are taken into account. Indeed, we aim at undertaking a more comprehensive study of this problem, relying on a Landau expansion of the free energy which seems not available at the moment \textbf{(CHECK)}, in a subsequent paper.
\section*{Acknowledgements}
S. T. is grateful to D. Chillingworth for introducing him to the subject of CT. This work was partly supported by the Italian GNFM (Gruppo Nazionale per la Fisica Matematica) \emph{Progetto Giovani}.


\begin{thebibliography}{99}

\bibitem{95dgpr}
{\sc P.G.~de Gennes and J.~Prost},
{\it The Physics of Liquid Crystals}, 
2nd Edition, Oxford Univ.\ Press, Oxford, 1995.

\bibitem{94virga}
{\sc E.G.~Virga},
{\it Variational theories for liquid crystals}, 
Chapman \& Hall/CRC, London, 1995.

\bibitem{80land}
{\sc L.D.~Landau and E.M.~Lifshitz},
{\it Course of Theoretical Physics: Statistical Physics}, 
Volume 5, Part I, 3rd Edition, Pergamon  Press, 1980.

\bibitem{93gilm}
{\sc R.~Gilmore},
{\it Catastrophe Theory for Scientists and Engineers},
Dover Publications, 1993. 
John Wiley \& Sons, 1981.

\bibitem{99dema}
{\sc M.~Demazure},
{\it Bifurcations and Catastrophes: Geometry of Solutions to Nonlinear Problems},
Springer-Verlag, Berlin, 1999.

\bibitem{93okad}
{\sc K.~Okada},
{\it Catastrophe Theory and Phase Transitions: Topological Aspects of Phase Transitions and Critical Phenomena},
Solid State Phenomena, 34 (1993), 1.

\bibitem{78golu}
{\sc M. Golubitsky},
{\it An Introduction to Catastrophe Theory and its Applications},
SIAM Review, 20, (1978), pp. 352--387.

\bibitem{73grwi}
{\sc R.~B. Griffiths and B. Widom},
{\it Multicomponent-fluid tricritical points},
Phys.\ Rev. A, 8 (1973), 2173.

\bibitem{73grif}
{\sc R.~B. Griffiths},
{\it Proposal for notation at tricritical points},
Phys.\ Rev. B, 7 (1973), 545.

\bibitem{sluck}
{\sc T. J. Sluckin and G. R. Luckhurst},
{Private communication} (2008).

\bibitem{03sovidu}
{\sc A.M.~Sonnet, E.G.~Virga and G.E.~Durand},
{\it Dielectric shape dispersion and biaxial transitions in nematic liquid crystals},
Phys.\ Rev.\ E, 67 (2003), 061701.

\bibitem{05dmvi}
{\sc G.~De Matteis and E. G. Virga},
{\it Tricritical points in biaxial liquid crystal phases},
Phys.\ Rev. E, 71 (2005), 061703.

\bibitem{05dmvi2}
{\sc G.~De Matteis and E. G. Virga},
{\it Criterion for tricritical points in liquid crystal phases},
in ``Progress in Nonlinear Differential Equations and Their Applications'',
Birkh\"{a}user, 68 (2006), pp. 55--74.

\bibitem{05dmrovi}
{\sc G. De Matteis, S. Romano and E. G. Virga},
{\it Bifurcation analysis and computer simulation of biaxial liquid crystals},
Phys.\ Rev. E, 72 (2005), 041706.

\bibitem{07dmbivi}
{\sc G. De Matteis, F. Bisi and E. G. Virga},
{\it Constrained stability for biaxial nematic phases},
Continuum Mech. Thermodyn., 19 (2007), pp. 1--23.


\bibitem{82huan}
{\sc C. C. Huang},
{\it The existence of a tricritical point in the smectic A--smectic C transition line},
Solid State Comm.,43 (1982), pp. 883--885.

\bibitem{87gahu}
{\sc C. W. Garland and M. E. Huster},
{\it Nematic--smectic-$C$ heat capacity near the nematic--smectic-$A$--smectic-$C$ point},
Phys. \ Rev. A, 35 (1987), 2365.

\bibitem{88shra}
{\sc S. Shashidhar, B. R. Ratna, Geetha G. Nair and S. Krishna Prasad},
{\it Mean-field to tricritical crossover behavior near the smectic-$A$--smectic-$C^{*}$ tricritical point},
Phys. \ Rev. Lett., 61 (1988), 547.

\bibitem{95brfu}
{\sc T. Br\"{a}uniger and B. M. Fung},
{\it Relation between the orientational ordering and the tricritical behavior for smectic-A to smectic-C phase transition},
J. Chem. Phys., 102 (1995), 7714.

\bibitem{08damu}
{\sc A. K. Das and P.K. Mukherjec},
{\it Tricritical behavior of the smectic-A to smectic-C phase transition in a liquid crystal mixture},
J. Chem. Phys., 128 (2008), 234907.

\bibitem{07bicate}
{\sc P. Biscari, M. C. Calderer and E. M. Terentjev},
{\it Landau-de Gennes theory of isotropic-nematic-smectic liquid crystal transitions},
Phys. \ Rev. E, 75 (2007), 051707.


\bibitem{74stra}
{\sc J. P. Straley},
{\it Ordered phases of a liquid of biaxial particles},
Phys. Rev. A, 10 (1974), 1881.

\bibitem{86long}
{\sc L. Longa},
{\it On the tricritical point of the nematic-smectic A phase transition in liquid crystals},
J. Chem. Phys., 85 (1986), pp. 2974--2985.

\bibitem{89long}
{\sc L. Longa},
{\it Order-parameter theories of phase diagrams for antiferroelectric smectic-A phases},
Liquid Crystals, 5 (1989), pp. 443--461.

\end{thebibliography}
\end{document}